# Multiscale Materials Modelling through Machine Learning – Hydrogen/Steel Interaction during Deformation

## M. Amir Siddiq[a,*]

[a]*School of Engineering, University of Aberdeen, Fraser Noble Building, AB24 3UE, Aberdeen, United Kingdom*

[*]Corresponding Author: amir.siddiq@abdn.ac.uk

## Abstract

This short paper presents the potential of using machine learning to predict materials behaviour in the context of hydrogen interaction with steel. Effort has been made to understand the quality, and amount of data needed to get improved predictions. An approach known as physics informed machine learning has been adapted in a simplified way through data classification to show the improvement in predictions. Proposed model eliminates the requirement to solve complex materials constitutive models and can work for any length scale, in the present case it is used for single crystalline steel interacting with steel under different types of loading.

## 1. Introduction

Since the birth of machine learning, it has evolved and applied in many areas of our daily life, such as recommendation and content personalisation systems on various multimedia (Netflix, YouTube, Disney); social media (Twitter, LinkedIn, Facebook); search engine (Google, Bing) platforms and so on. Machine learning algorithms are data-hungry and require massive amount of data for reliable predictions. Data can be in different forms such as numerals, words, pictures, clicks and anything relevant which can be accessed. Materials science is no exception and in the last few years machine learning is used in this area to shift from knowledge-driven [1]–[3] to data-driven approaches [4]. Machine learning in materials science is mostly used as a black box and hence can pose several challenges. These challenges are associated with the quantity, format, and availability of reliable experimental data. Data in the context of materials science could be microstructure, deformation curves, failure behaviour, interaction of the materials with the environment, and so on. The question here is What type of data is necessary and How much data is needed for confident and reliable predictions?

The answer is not straight forward and depends on many factors including what needs to be predicted. In the following, an example of hydrogen interaction with a single crystal austenitic steel is presented to demonstrate the importance of the type and quantity of data needed for reliable predictions using machine learning.

## 2. Predicting Hydrogen, Single Crystal Steel, and Loading Interaction
### 2.1 Computational Modelling through Mechanistic Crystal Plasticity Theory

Hydrogen storage and transportation devices use different materials including steels. As shown in Figure 1 (top left), hydrogen can diffuse in steel and interact with defects (cavities), precipitates, vacancies, grain boundaries, dislocations, and interphases. The presented example is focussed on hydrogen interaction with cavities (voids) during deformation. Dataset used for machine learning is generated through a mechanistic model [5], [6] after experimental validations. Mechanistic model was based in crystal plasticity theory [3], [7] implemented in finite element methods (CPFEM). CPFEM was then used to investigate the interaction of hydrogen with single crystal deformation, defects and loading type, where plastic deformation was governed by dislocation motion on 12 possible slip systems in face-centred-cubic (FCC) crystal of austenitic stainless steel. While hydrogen enhanced localised plasticity (HELP) mechanism was incorporated in single crystal plasticity theory to account for hydrogen effects (for discussion see [5], [6]). It was found that the stress strain curves for different initial crystal orientation are a strong function of hydrogen concentration. It was also reported that initial crystal orientation relative to applied displacement is found to affect the influence of hydrogen on plastic deformation and void growth. The distribution and magnitude of hydrogen in traps, especially in the vicinity of the void, was also found to have a strong dependence on initial hydrogen concentration (see bottom left contour plot in Figure 1). The data generated through CPFEM contains single crystalline information during deformation; this includes stresses, strains, initial crystal orientations, initial defect sizes, initial hydrogen concentrations, plastic slip activity, triaxialities, lode parameters, along with evolutions of these quantities with deformation (see bottom left contour plot in Figure 1 showing interaction of hydrogen in traps around the voids during deformation).

### 2.2 Predicting through Machine Learning

Although mechanistic model discussed in the previous section accounts for the physics of the problem, it requires to solve a number of complex differential equations which require a large number of material parameters to be identified through experiments. There have been investigations to deal with these issues, such as by developing automatic parameter identification procedure [8], [9], or using data driven finite element methods [4], [10].

The data generated (Figure 1 middle part) in section 2.1 in the form of material data repository is fed into the machine learning (support vector machines) through coding in Python. Support vector machines (SVM) tries to find a line/hyperplane in multidimensional space that separates difference classes of data. Unlike other regression models that try to minimise the error between the real and predicted values, the support vector regression (SVR) tries to fit the best line within a threshold value (distance between hyperplane and boundary line). In a sense, SVR is like linear regression in that the equation of the line is y=wx+b, while in SVR, this straight line is referred to as hyperplane. The data points on either side of the hyperplane that are closest to the hyperplane are called support vectors which are used to plot the boundary line. Initially 60% of the data is used to train the model, while 40% data is used for predictions. It can be seen in the Figure 1 (case 1), machine learning (SVM) struggles to predict the stress-strain response for different loading and hydrogen conditions. After careful quantification of this uncertainty, dataset was classified into elastic (linear) and plastic (nonlinear) parts of the single crystal deformation using SVM. And then through selective training data, all the elastic data was used as part of the training while rest of the plastic deformation data which constituted 60-40 proportion of training and prediction data. Results of this modification are plotted as case 2 in the Figure 1. It can be inferred from the contour plots that the predictions obtained from physics informed machine learning compared well with CPFEM results for equivalent stress, strain and triaxiality while using the same type and amount of data.

## 3. Conclusions

In the presented work a supervised machine learning algorithm, support vector machines, is used to learn and predict the hydrogen interaction with steel at single crystalline level. The trained model doesn't require any complex differential equation to describe the physics. Results show a very good agreement with the dataset. It is also found that using machine learning algorithms blindly can bring a number of uncertainties, one of it has been demonstrated and removed through physics informed machine learning via selective training through data analytics.

Looking forward, there is no doubt that machine learning will play a crucial role in predicting materials science, however the question remain subtle, especially what type of data is necessary and most importantly how much data is needed? One option is discussed above, viz development of physics informed machine learning algorithms to remove experimental uncertainty and data starvation issues, this is the top trend in machine learning based materials science.

### Acknowledgements
No external funding was received for this project.

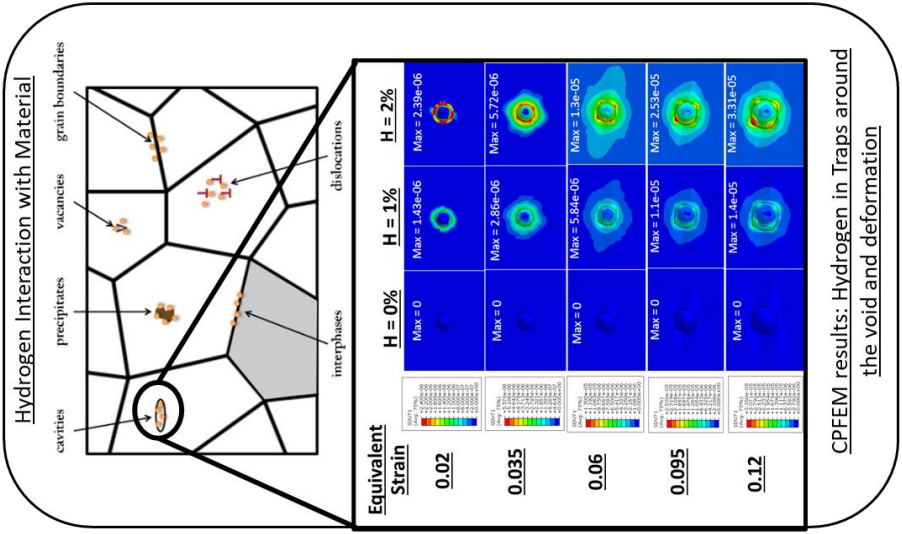
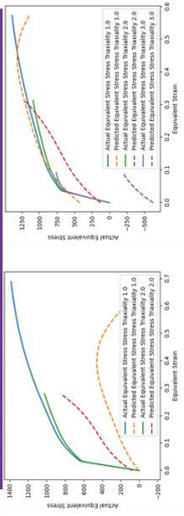
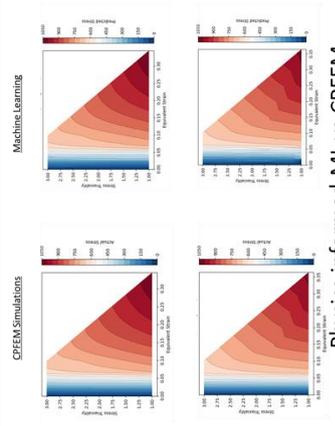

**Figure 1: Machine learning and predicting hydrogen interaction with single crystal steel during deformation**